\documentclass{elsart}
\usepackage{amssymb,epsfig}
\begin{document}
\begin{frontmatter}

\title{Effect of Hetrovalent substitution at Mn site on the Magnetic and Transport Properties of La$_{0.67}$Sr$_{0.33}$MnO$_3$}
\author[gu]{K. R. Priolkar\corauthref{krp}}\ead{krp@unigoa.ac.in} and
\author[iuc]{R. Rawat} 
\corauth[krp]{Corresponding author}
\address[gu]{Department of Physics, Goa University, Taleigao-Plateau, Goa, India 403 206}
\address[iuc]{UGC-DAE Consortium for Scientific Research, University Campus, Khandwa Road, Indore, India 452 017}

\begin{abstract}
Magnetic and transport properties of Ti substituted La$_{0.67}$Sr$_{0.33}$MnO$_3$ are drastically affected with a change in preparation conditions. Low temperature infra-red absorption measurements reveal that this is perhaps due to inhomogeniety in substitution of Ti$^{4+}$ on Mn sites. It is found that, in the high temperature annealed samples, the substitution of Ti supresses the double exchange interaction due to the formation of Mn$^{3+}$-O-Ti$^{4+}$ chains. While in the low temperature annealed case substitution of Ti causes formation of isolated ferromagnetic clusters linked to each other by a variable range hopping polaron.  
\end{abstract}


\end{frontmatter}

\section{Introduction}
In recent years, the hole doped rare-earth manganites are one of the most intensively studied materials, being strongly attractive to both basic research and technology due to their rich physics and potential prospects of application \cite{good,mart,cnr}. Most of the undoped compounds (for example, LaMnO$_3$) are insulators with antiferromagnetic order. Replacing rare-earth by a alkaline earth metals like Ca, Sr or Ba results in conversion of an appropriate number of Mn$^{+3}$ to Mn$^{+4}$. This gives rise to ferromagnetic Mn$^{+3}$-O-Mn$^{+4}$ double exchange interactions \cite{zen}. In this interaction the e$_g$ electron from Mn$^{+3}$ is transferred to Mn$^{+4}$ with a parallel spin configuration. In pertinent substitution ranges, where double exchange interactions dominate, some compounds can be metallic ferromagnets and exhibit colossal magnetoresistance at T$_c$. Jahn-Teller type electron-phonon interactions are greatly responsible for the appearance of such properties \cite{mil1,mil2,hwang}. Substitution especially at Mn site strongly affects the character of Jahn Teller polarons and therefore the magnetic and transport properties mainly due to difference in size of the dopant \cite{hwang,teresa} i.e., on the degree of GdFeO$_3$-type lattice distortion and lattice disorder. The disorder induced by dopant ions occupying heterovalent sites has not been studied so far. Such a disorder can have tremendous effect on the properties of these materials. Ti doped CMR manganites offer such an opportunity.  Ionic radius of Ti$^{4+}$ ion is known to be in between the ionic radii of Mn$^{4+}$ and Mn$^{3+}$. Therefore there exists a distinct possibility that a fraction of Ti$^{4+}$ ions can occupy Mn$^{3+}$ sites in addition to the Mn$^{4+}$ sites. The available literature on Ti substitution in manganites  indicates that Ti$^{4+}$ ions occupy the isovalent Mn sites in these materials. \cite{sah,hu,troy,kal,liu,kim,uly1,uly2,alva,nam} Recent reports have shown that though it prefers to occupy Mn$^{4+}$ sites, at high doping levels it also occupies Mn$^{3+}$ sites. \cite{xx}. In order to see the effect of Ti$^{4+}$ occupying both Mn$^{3+}$ and Mn$^{4+}$ sites two sets of compounds having compositions, La$_{0.67}$Sr$_{0.33}$Mn$_{1-x}$Ti$_x$O$_3$ with $x$ = 0, 0.1, 0.2 \& 0.33 were prepared employing different preparation procedures. Both the sets of compounds have been studied by X-ray diffraction, resistivity, susceptibility, magnetoresistance and infra red spectroscopy. It is found that magnetic and transport properties of the two sets are seriously affected and could be linked to formation of a Jahn teller polarons in these compounds.
 
\section{Experimental}
Two batches of samples of the type La$_{0.67}$Sr$_{0.33}$Mn$_{1-x}$Ti$_{x}$O$_3$ with $x$ = 0, 0.1, 0.2 and 0.33 have been prepared by solid state reaction method. In case of first batch the samples were well ground for over 4 hours and annealed at 1300$^\circ$C several times to obtain well ordered samples. These samples are referred to as HT samples. While in the other case the samples were ground for a much lesser time $\sim$ 30 minutes and annealed at 1100$^\circ$C. These were termed as LT samples. All the samples were characterized by X-ray diffraction using Cu K$_\alpha$ radiation. Resistivity as a function of temperature was measured in the temperature range 80K to 350K using the standard four-probe setup. DC susceptibility was measured using a Faraday Balance in the temperature range 15K to 300K and a field of 100Oe. Infield resistivity measurements were carried out in logitudinal geometry down to 30 K and magnetic fields up to 5 T using OXFORD Spectromag 10 T superconducting magnet. Infrared measurements were performed on Shimadzu FTIR-8900 spectrophotometer at room temperature in transmission mode in the range of 350 cm$^{-1}$ to 1000 cm$^{-1}$. The samples were mixed with KBr in the ratio 1:100 by weight and pressed into pellets for IR measurements. Temperature variation was achieved using an Oxford Optical cryostat with KRS5 windows. 
 
\section{Results and Discussion}
\begin{figure}[h]
  \centering
  \epsfig{file=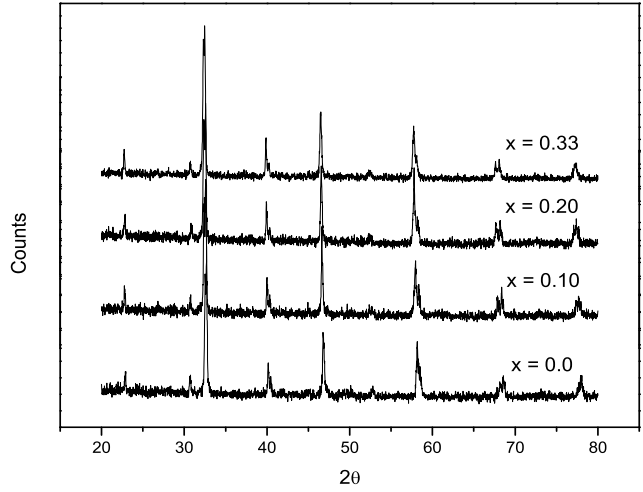, width=10cm, height=8cm}
  \caption{X-ray diffraction patterns for La$_{0.67}$Sr$_{0.33}$Mn$_{1-x}$Ti$_{x}$O$_3$ for HT samples}
  \label{fig:xrd-o}
\end{figure}

\begin{figure}
\centering
 \epsfig{file=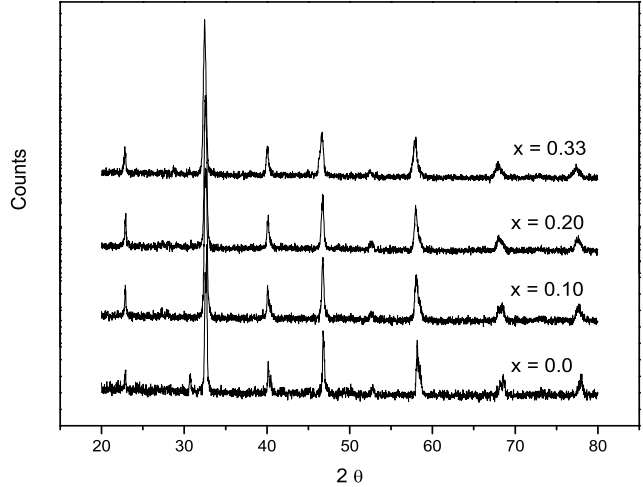, width=10cm, height=8cm}
\caption{X-ray diffraction patterns for La$_{0.67}$Sr$_{0.33}$Mn$_{1-x}$Ti$_{x}$O$_3$ for LT samples}
\label{fig:xrd-d}
\end{figure}

The powder X-ray diffraction (XRD) plots presented in Fig. \ref{fig:xrd-o} and \ref{fig:xrd-d} show that the compounds in the entire substitution range have rhombohedral $R\bar3c$ structure. Rietveld refinement of the XRD plots was performed in the hexagonal setting of the $R\bar3c$ space group, in which La atoms occupy 6a (0,0,$1\over4$), Mn  6b (0,0,0) and O 18e(x,0,$1\over4$) positions. The obtained values of lattice parameters are presented in Table \ref{tab:3tab1}. It can be seen from the table that for the HT series the lattice parameters increase steadily with increasing Ti concentration. Such an increase has also been seen by others in these compounds \cite{kal,nam}. The increase in the lattice parameters can be related to the larger ionic radii of Ti$^{4+}$ ion as compared to that of Mn$^{4+}$ ion ($r_{Ti^{4+}}$ = 0.605\AA, and $r_{Mn^{4+}}$ = 0.54\AA). The Mn(Ti)-O bond distances calculated from the structural parameters also increases with $x$ but the Mn(Ti)-O-Mn(Ti) bond angles change only slightly with $x$. This indicates that each BO$_6$ octahedron undergoes only a little distortion with Ti substitution. In the case of LT samples the lattice parameters remain nearly constant throughout the series.
The calculation of bond distances and bond angles shows that both of them remain constant throughout the series. 

\begin{table}[h]
\caption{Lattice parameters for La$_{0.67}$Sr$_{0.33}$Mn$_{1-x}$Ti$_{x}$O$_3$ 
 (x=0, 0.1, 0.2 \& 0.33) for HT and LT samples. Figures in the brackets indicate uncertainty in the last digit.}
\vspace{0.2cm}
\centering
\begin{tabular}{|c|c|c|c|c|}
\hline
lattice constants & x = 0 & x = 0.1 & x = 0.2 & x = 0.33\\
 
\hline
a(HT) & 5.498(7) & 5.520(7) & 5.530(6) & 5.536(6) \\ 
c(HT) & 13.367(5) & 13.400(6) & 13.422(5) & 13.452(4)\\
a(LT) &     & 5.509(8) & 5.511(9) & 5.514(8) \\
c(LT) &     & 13.372(8) & 13.383(7) & 13.391(9) \\
\hline 
\end{tabular}
\label{tab:3tab1}
\end{table}

\begin{figure}[h]
\centering
\epsfig{file=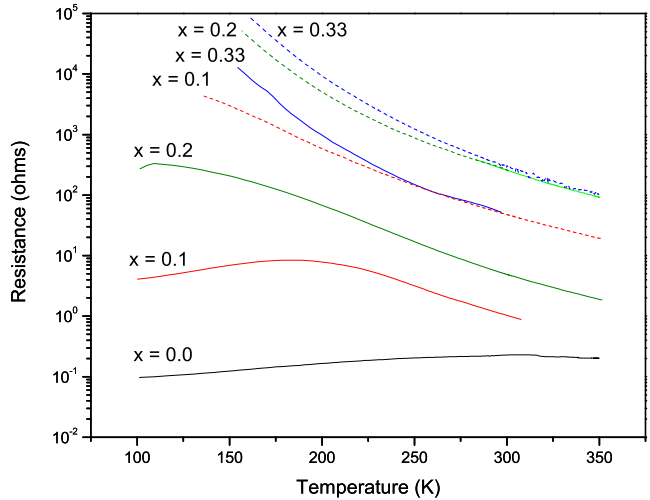, width=10cm, height=8cm}
\caption{Temperature dependence of the electrical resistivity for samples La$_{0.67}$Sr$_{0.33}$Mn$_{1-x}$Ti$_{x}$O$_3$.The solid lines depict the resistivity of HT compounds while the dashed ones of LT compounds.}
\label{fig:res}
\end{figure}

Figure \ref{fig:res} shows the resistivity plots of all the compounds as a function of temperature. In the case of  x = 0, a metal-insulator transition is observed at about 320K. This temperature is lower than the transition temperature reported for this compound in literature and could be due to polycrystalline nature of the samples reported here. 
 
Likewise compounds with x = 0.1 and x = 0.2 from the HT set exhibit insulator to metal transitions but at lower temperatures. Magnitude of resistivity is also found to increase with increasing Ti concentration. In the case of  x = 0.33 sample although the resistance increases by several orders of magnitude, no metal to insulator transition is observed. A similar trend and transition temperatures are also observed in Ti substituted La$_{1-x}$Sr$_x$MnO$_3$ reported by \cite{kal,kim,nam}. The decrease in the metal to insulator transition temperature $T_{MI}$ and increase in the resistivity with increasing Ti doping level can be ascribed to the replacement of some of the Mn$^{3+}$-O-Mn$^{4+}$ bonds by the Mn$^{3+}$-O-Ti$^{4+}$ bonds. The higher radius of Ti leads to an increase of the average (Mn,Ti)-O distance that suppresses double exchange interaction between mixed valent Mn ions and hence reduces the ferromagnetic coupling between neighbouring manganese. It is also clear that replacing Mn by Ti hampers the electron transfer through the Mn$^{+3}$ - O$^{-2}$ - Mn$^{+4}$ network seriously. 

In case of LT samples, no insulator to metal transition is observed in resistivity. The magnitude of resistivity as well, is much higher than their HT counterparts. This could be either due to  grain boundary effect or the disorder in Ti doping at lower annealing temperatures. The ionic size of Ti$^{4+}$ being in between ionic sizes of Mn$^{3+}$ and Mn$^{4+}$, there is a possibility of formation of  Mn$^{4+}$-O-Ti$^{4+}$ chains in addition to Mn$^{3+}$-O-Ti$^{4+}$. Recently, in the case of La$_{0.7}$Sr$_{0.3}$Mn$_{1-x}$Ti$_x$O$_3$ type thin films, it was found that with an increase in Ti content, the lattice constant first increased due to the substitution of Mn$^{4+}$ with Ti$^{4+}$, and for $x \ge 0.3$ it tended to decrease due to the substitution of Mn$^{3+}$ by Ti$^{4+}$ with a larger ionic radius, as well as when excess oxygen was introduced \cite{xx}. 

\begin{figure}[h]
\centering
\epsfig{file=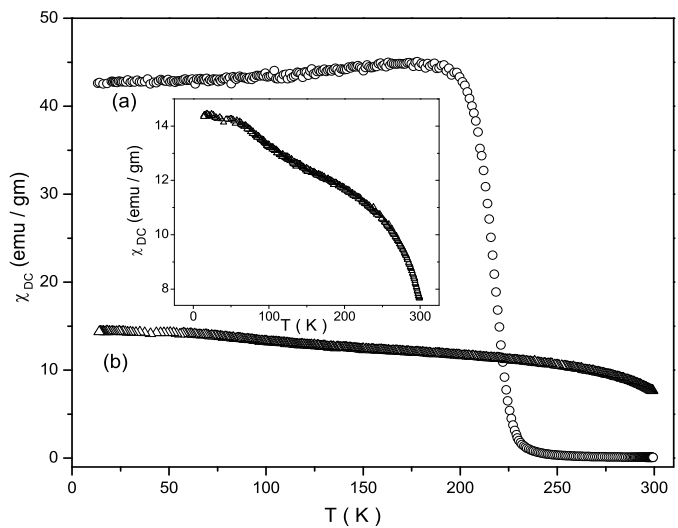, width=10cm, height=8cm}
\caption{Susceptibility as a function of temperature in case of x = 0.1 HT (a) and LT (b) samples}
\label{fig:sus}
\end{figure}

\begin{figure}[h]
\centering
\epsfig{file=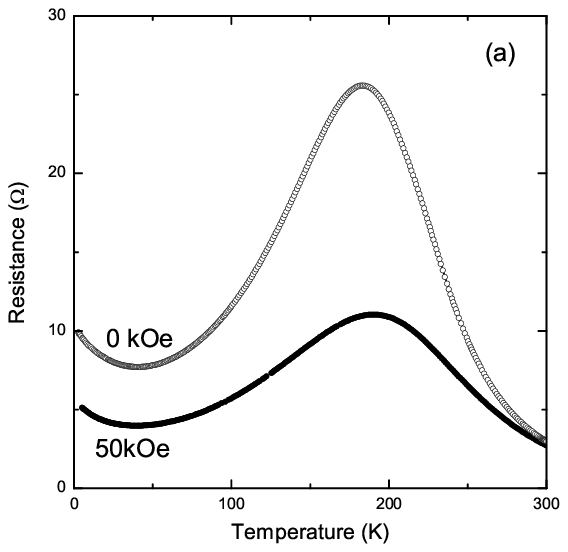, width=6cm, height=8cm}
\epsfig{file=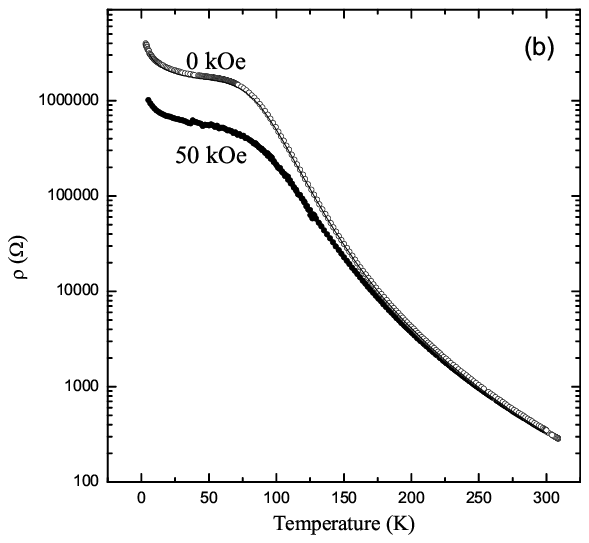, width=6cm, height=8cm}
\caption{Magnetoresistance as a function of temperature in case of x = 0.1 HT (a) and LT (b) samples}
\label{fig:mr}
\end{figure}

\begin{figure}[h]
\centering
\epsfig{file=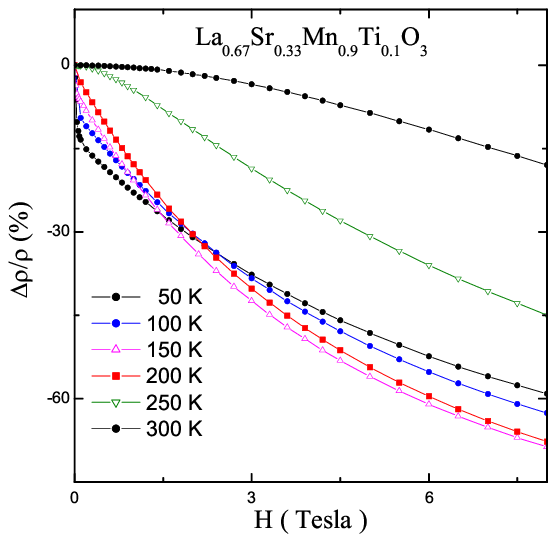, width=6cm, height=8cm}
\epsfig{file=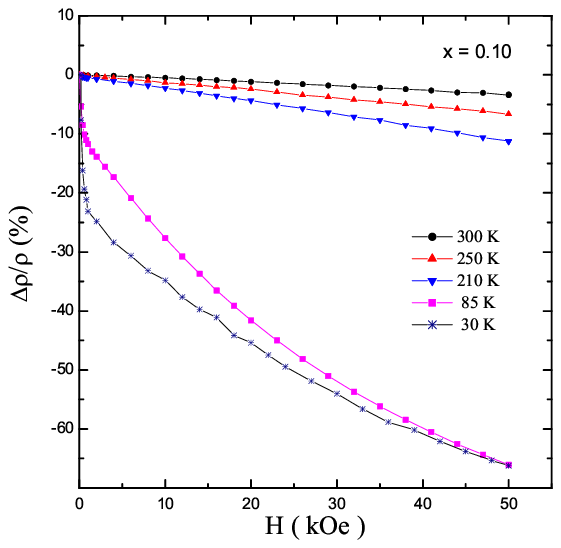, width=6cm, height=8cm}
\caption{Magnetoresistance as a function of field in case of x = 0.1 HT (a) and LT (b) samples}
\label{fig:mr1}
\end{figure}

Both the above scenario should have its effect on other physical properties as well. In order to check this a comparison of susceptibility and magnetoresistance of $x$ = 0.1 sample from both HT and LT groups was made. Figure \ref{fig:sus} presents susceptibility as a function of temperature for the two samples. A clear paramagnetic to ferromagnetic transition is seen in the HT sample at a temperature close to that of $T_{MI}$. However, in case of LT compound, there seems to be a possibility of a weak ferromagnetic like transition just above 300K. While the ferromagnetic transition in the HT sample can be attributed to DE mechanism, no insulator to metal transition is present in the LT sample.  
Furthermore, magnetoresistance (MR) signal of about 60\% at T$_{MI}$ in a field of 5T is seen in the case of HT sample while for the LT sample the MR at 300K (temperature at which a weak ferromagnetic transition may be present) is only about 10\% (see figure\ref{fig:mr}). The MR of LT sample does reach a value of 60\% at about 85K. At about the same temperature the susceptibility also shows signatures of second magnetic transition as can be seen more clearly in the inset of figure \ref{fig:sus}. If grain boundaries are the cause of higher resistance in the case of LT samples, then the volume fraction sensitive properties like magnetization should not be affected \cite{hwang}. However, it can be clearly seen that the behaviour of magnetization for $x$ = 0.1 of both HT and LT samples is different. Furthermore, studies \cite{hwang,klein,kum,gupta,li} examining the effect of grain boundaries on magnetotransport properties of manganite thin films have shown that low field magnetoresistance increases sharply due to intergrain tunneling. In figure \ref{fig:mr1} isothermal magnetoresistance as a function of applied field at various temperatures is presented. It can be seen that grain boundary effects are not seen at higher temperatures where the magnetic transition occurs in LT sample.  At lower temperatures ($<$ 100K) grain boundary effects are seen and for both the samples, low field MR increases with decreasing temperature. Therefore, the ferromagnetic transition at $\sim$ 300K in the LT sample can be ascribed to a hopping polaron that connects isolated Mn rich regions ferromagnetically. It may be mentioned here that the behavior of resistivity in the insulator region of both these samples can be explained by Mott's variable range hopping model.

\begin{figure}
\centering
\epsfig{file=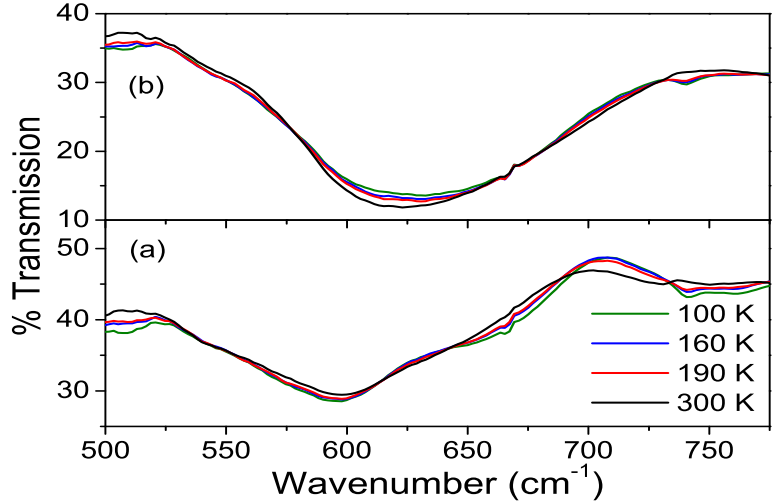, width=12cm, height=8cm}
\caption{Plot of infra-read transmission at different temperatures in case of x = 0.1 HT (a)  and LT (b) samples.}
\label{fig:ir}
\end{figure}

Insulator to metal transition in manganites that is correlated with the paramagnetic to ferromagnetic transition, is essentially due to competition between the DE interaction that induces delocalization and the strong Jahn-Teller coupling that tends to localize the carrier in a local lattice distortion, viz. the formation of lattice polarons \cite{mill,roder}. To understand the differences in physical properties of HT and LT samples, infra-red (IR) absorption measurements as a function of temperature have been carried out on $x$= 0.1 and 0.2 samples. IR spectra were recorded in the temperature range 80K $\le$ T $\le$ 300K such that they span the metallic as well as the insulating region. The spectra in the frequency range 500 to 750 cm$^{-1}$ for $x$ = 0.1 sample are presented in figure \ref{fig:ir}. It can be clearly seen that as the HT sample undergoes a insulator to metal transition a new mode of vibration as absorption dip is observed  at about 650cm$^{-1}$ in the 100K spectra. This vibration mode is in addition to the main absorption dip at 600cm$^{-1}$ which is due to Mn-O bond stretching. 

It has been reported in literature, especially in case of La$_{0.7}$Ca$_{0.3}$MnO$_3$  that across the  metal insulator transition the Mn-O stretching mode hardens by about 30cm${^-1}$, increases in width and develops a fano type symmetry due to interaction of phonons with changing polaronic background \cite{kim1}. However, in the Ti doped HT sample the Mn-O stretching mode at about 600cm$^{-1}$ is present even in the metallic state along with a new mode at about 650 cm$^{-1}$. It may be noted here that this new mode is absent in the LT sample as can be seen from figure \ref{fig:ir}(b). Therefore this new peak is connected to the appearance of metal-insulator transition in the samples. Substitution of Ti$^{+4}$ for Mn results in Mn$^{+3}$-O$^{-2}$-Ti$^{+4}$ type chains along with Mn$^{+3}$-O$^{-2}$-Mn$^{+4}$ DE pairs. Therefore only those Mn$^{3+}$ ions participating in DE interaction will interact with the polaronic background resulting in a shift in the stretching mode. As there are Mn$^{3+}$ ions bonded also to Ti$^{4+}$ there will also be a unshifted stretching mode at 600cm$^{-1}$. 

\begin{figure}
\centering
\epsfig{file=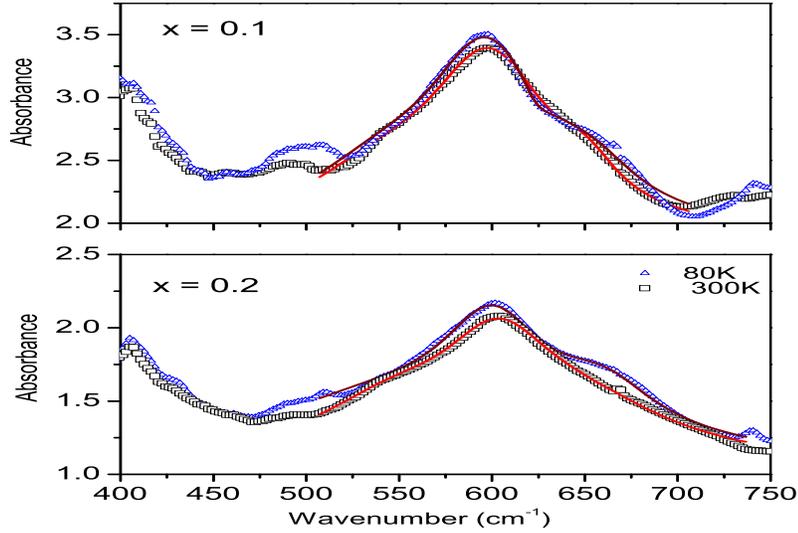, width=12cm, height=8cm}
\caption{Infra-red absorption at 300K and 80K in case of x = 0.1 and 0.2 (HT) samples. The solid line is a fit to data in the range 500 to 740 cm$^{-1}$.}
\label{fig:irfit}
\end{figure}

In SrTiO$_3$, Ti-O stretching modes appear at  540cm$^{-1}$ and 670cm$^{-1}$ \cite{liv}. Therefore it is possible that the actual position of the hardened mode is lower than it appears in figure \ref{fig:ir}(a). In order to ascertain the exact position of this mode the IR spectra in the region 500 - 740 cm$^{-1}$ at 300 K and 100 K were fitted with three and four Lorentzian's respectively in the case of $x$ = 0.1 and 0.2 HT samples. The results are presented in figure \ref{fig:irfit} The positions of the Mn-O (600cm$^{-1}$) and Ti-O (540 and 670cm$^{-1}$) stretching frequencies were fixed to their known values. While the 300K spectra could be fitted well with these three modes (1 Mn-O and 2 Ti-O), the spectra at 100K could only be fitted with an additional mode at 624 cm$^{-1}$. This position of the shifted mode agrees very well with the results obtained in case of La$_{0.7}$Ca$_{0.3}$MnO$_3$ \cite{kim1}. The new mode is therefore due to Jahn Teller polarons associated with Mn$^{3+}$-O stretching. 

In the case of LT samples, as can be seen from figure \ref{fig:ir}(b) the 600cm$^{-1}$ mode is a quite broadened. This can be related to the presence of disorder in the Ti doping. In this sample Ti$^{4+}$ substitution causes regions  wherein double exchange is active and regions which are devoid of double exchange pairs. Perhaps the two regions are in such a proportion that the strengths of respective stretching modes are nearly same resulting in a broad absorption dip. 

The Mn rich regions whrein double exchange is active, can order ferromagnetically. However, these regions are separated by Ti rich regions which isolate them so as not to exhibit either insulator to metal transition or magnetoresistance at the T$_C$. Isolated Mn rich regions may be linked to each other by a variable hopping range polaron. Such a linkage could be the cause of weak ferromagnetism in these samples.

\section{Conclusions}
In this paper results of transport, magnetic and optical properties of Ti doped La$_{0.67}$Sr$_{0.33}$MnO$_3$ manganites annealed at two different temperatures are presented. It is seen that low temperature annealing leads to vastly different properties like higher ferromagnetic transition temperature, absence if metal-insulator transition, absence of MR etc. Grain boundary contribution alone cannot explain all the results. From the magnetic, transport and spectroscopic measurements it is concluded that in the high temperature annealed samples the Ti$^{4+}$ ion replaces Mn ions leading to formation of Mn$^{3+}$-O-Ti$^{4+}$ bonds. While in the low temperature annealed case substitution of Ti leads to formation of inhomogeneous short range ordered ferromagnetic clusters. The weak ferromagnetic signal observed at 300K in the LT samples could then be due to the isolated DE pairs connected via a variable range hopping polaron. 

\section*{Acknowledgments}
KRP would like to thank Department of Science and Technology, Government of India for financial assistance under the project No. SR/FTP/PS-19/2003. Help of Ms. Shahin Desai and Ms. Bhakti Bandekar in sample preparation is also acknowledged.


\begin{thebibliography}{50}
\bibitem{good} J. B. Goodenough, Rep. Prog. Phys. {\bf 67} (2004) 1915.
\bibitem{mart} M. B. Soloman and M. Jaime, Rev. Mod. Phys. {\bf 73} (2001) 583.
\bibitem{cnr} C. N. R. Rao, A. Arulraj, A. Cheetham and B. Raveau, J. Phys.: Condens. Matter {\bf 12} (2001) R83
\bibitem{zen} C. Zener, Phys. Rev., {\bf 82} (1951) 403.
\bibitem{mil1} A. J. Millis, P. B. Littlewood and B. I. Shraiman, Phys. Rev. Lett. {\bf 74} (1995) 5144.
\bibitem{mil2} A. J. Millis, B. I. Shraiman and R. Mueller, Phys. Rev. Lett. {\bf 77} (1996) 175.
\bibitem{hwang}H. Y. Hwang, S. W. Cheong, P. G. Radaelli, M. Marezio and B. Battlog, Phys. Rev. Lett. {\bf 75} (1995) 914.
\bibitem{teresa} J. M. de Teresa, M. R. Ibarra, J. Garcia, J. Blasco, C. Ritter, P. A. Algarabel, C. Marquina and A. del Moral, Phys. Rev. Lett. {\bf 76} (1996)3392.
\bibitem{sah}M. Sahana, A. Venimadhav, M. S. Hegde, K. Nenkov, U. K. Ro$\beta$ler, K. Dorr, K. -H. Muller, J. Magn. Magn. Mater. {\bf 260} (2003) 361.
\bibitem{hu} J. Hu, H. Qin, J. Chen, Z. Wang, Mat. Sci. Engg. B {\bf 90} (2002) 146.
\bibitem{troy} I. O. Troyanchuk, M. V. Bushinsky, H. Szymczak, K. Barner and A. Maignan, Eur. J. Biochem. {\bf 28} (2002) 75.
\bibitem{kal} N. Kallel, G. Dezanneau, J. Dhahri, M. Oumezzine and H. Vincent, J. Magn. Magn. Mater. {\bf 261} (2003) 56.
\bibitem{liu} Y -H. Liu, B -X. Huang, R -Z Zhang, X -B Yuan, C -J. Wang and L -M Mei, J. Magn. Magn. Mater. {\bf 269} (2004) 398.
\bibitem{kim} M. S. Kim, J. B. Yang, Q. Cai, X. D. Zhou, W. J. James, W. B. Yelon, P. E. Parris, D. Buddhikot and S. K. Malik, Phys. Rev. B {\bf 71} (2005) 014433.
\bibitem{uly1} A. N. Ulyanov, Y -M. Kang, S -I. Yoo, D -S. Yang, H. M. Park, K -W. Lee and S -C. Yu, J. Magn. Magn. Mater. {\bf 304} (2006) e331.
\bibitem{uly2} A. N. Uluanov, D -S. Yang, K -W. Lee, J -M. Greneche, N. Chau and S -C. Yu, J. Magn. Magn. Mater. {\bf 300} (2006) 175. 
\bibitem{alva} I. Alvarez-Serrano, M. L. Lopez, C. Pico and M. L. Viega, Sol. State Sci. {\bf 8} (2006) 37.
\bibitem{nam} D. N. H. Nam, L. V. Bau, N. V. Khiem, N. V. Dai, L. V. Hong, N. X. Phuc, R. S. Newrock and P. Nordblad, Phys. Rev. B {\bf 73} (2006) 184430.
\bibitem{xx} X. B. Zhu, Y. P. Sun, R. Ang, B. C. Zhao and W. H. Song, J. Phys D: Appl. Phys. {\bf 39} (2006) 625.
\bibitem{hwang} H. Hwang, S. -W. Cheong, N. P. Ong and B. Batlogg, Phys. Rev. Lett. {\bf 77} (1996) 2041.
\bibitem{klein} J. Klein, C. Hoffnerm S. Uhlenbruck, L. Alff, B. Buchner and R. Gross, Europhys. Lett., {\bf 43} (1999) 371. 
\bibitem{kum} N. D. Mathur, G. Burnell, S. P. Isaac, T. J. Jackson, B. -S. Teom J. L. MacManus-Driscoll, L. F. Cohen, J. E. Evetts and M. G. Blamire, Nature {\bf 387} 266 (1997).
\bibitem{gupta} A. Gupta, Phys. Rev. B {\bf 54} (1996) R15629.
\bibitem{li} K. Steenbeck, T. Eick, K. Kirsch, K. O'Donnell and E. Steinbeiss, Appl. Phys. Lett., {\bf 71} (1997) 968.
\bibitem{mill} A. J. Millis, Nature {\bf 392} (1998) 147.
\bibitem{roder} H. Roder and A. R. Bishop, Phys. Rev. Lett., {\bf 76} (1996) 1356.
\bibitem{kim1}K. H. Kim, J. Y. Gu, H. S. Choi, G. W. Park and T. H. Noh, Phys. Rev. Lett., {\bf 77} (1996) 1877.
\bibitem{liv} R. A. Nquist and R. O. Kagel, {\it ``Infrared Spectra of Inorganic Compounds''} (1971) Academic Press NY. 
\end{thebibliography}
\end{document}